\documentclass[iop]{emulateapj}
\usepackage{mathptmx}

\begin{document}

\title{Measurement of The Broad Line Region Size in a Luminous MACHO Quasar}
\author{Doron Chelouche\altaffilmark{1}, Eliran Daniel\altaffilmark{2}, \& Shai Kaspi\altaffilmark{2,3}}
\altaffiltext{1} {Department of Physics, Faculty of Natural Sciences, University of Haifa, Haifa 31905, Israel; doron@sci.haifa.ac.il}
\altaffiltext{2}{School of Physics \& Astronomy and the Wise Observatory, Tel-Aviv
University, Tel-Aviv 69978, Israel; shai@wise.tau.ac.il, elirandviv@gmail.com}
\altaffiltext{3} {Department of Physics, Technion, Haifa 32000, Israel}
\shortauthors{Chelouche et al.}
\shorttitle{The BLR size of a Luminous MACHO Quasar}

\begin{abstract}

We measure the broad emission line region (BLR) size of a luminous, $L\sim 10^{47}\,{\rm erg\,s^{-1}}$, high-$z$ quasar using broadband photometric reverberation mapping. To this end, we analyze $\sim$7.5 years of photometric data for MACHO\,13.6805.324 ($z\simeq1.72$) in the $B$ and $R$ MACHO bands and find a time delay of $180\pm40$\,days in the rest frame of the object.  Given the spectral-variability properties of high-$z$ quasars, we associate this lag with the rest-UV iron emission blends. Our findings are consistent with a simple extrapolation of the BLR size-luminosity relation in local active galactic nuclei to the more luminous, high-$z$ quasar population. Long-term spectroscopic monitoring of MACHO\,13.6805.324 may be able to directly measure the line-to-continuum time-delay and test our findings.

\end{abstract}


\section{Introduction}

Our understanding of galaxy and black hole (BH) formation and co-evolution has progressed significantly in recent years \citep{wan99,fer00,har04,pen06,ben10,cis11} with new observations being able to probe the first epoch of quasar activity, and place interesting constraints on the mechanisms responsible for BH growth and galaxy formation \citep{hu03,tra11}. Gaining further physical insight requires that the BH masses be determined with good accuracy in a large sample of objects. 

Presently, the best means for weighing BHs in quasars is via the reverberation mapping (RM) technique \citep{pet93}, which measures the size of the broad line region (BLR). This, combined with a measure of the velocity dispersion of the BLR, can be used to estimate the BH mass \citep{pet04}. To date, BH masses in $\sim 45$ low-$z$ objects have been measured in this way leading to various scaling laws with other quasar properties. These relations are often extrapolated to high-$z$ objects to allow for the {\it indirect} estimate of their BH mass \citep{net03}. Nevertheless, it is not clear that such extrapolations are meaningful, and a more direct measure of the BH mass in such objects  is highly desirable. 

Here we report new measurements for the BLR size in a $z\simeq 1.72$ luminous quasar, using photometric RM \citep{cd11}, and by analyzing $\sim7.5$ years of data from the MACHO survey \citep{geh03}. Section 2 presents the analysis and results, with a follow-up discussion in \S3.

\section{Analysis \& Results}

To date, some 200 quasars have been confirmed behind the Magellanic clouds \citep{dob02,geh03,koz11}. In the course of analyzing their photometric data, we report our findings for MACHO\,13.6805.324, an $m_R=18.66$ quasar at $z\simeq 1.72$ having a monochromatic luminosity\footnote{To estimate the monochromatic luminosity of MACHO\,13.6805.324 at 1350\AA, we take the reported $R$-magnitudes from \citet{geh03}, employ a standard K-correction, and account for the extinction behind the LMC using the results of  \citet[see: http://ngala.as.arizona.edu/dennis/lmcext.html]{zar04}. To this end, we  average over $50$ stars of all types in the direction of the quasar, and identify the upper turn-over in the extinction distribution, at $A_V\simeq 1.3$\,mag, with the effective extinction of the background source. We conservatively estimate the uncertainty on the luminosity to be a factor 3, to account for the potentially patchy structure of LMC disk. Our estimate is consistent with $m_B$-$m_R$ for MACHO\,13.6805.324 being larger by $\sim 0.67$\,mag than typical, as concluded by integrating the \citet{dvb01} composite quasar spectrum over the MACHO bands, and assuming a Galactic extinction law. Using concordance (0.3,0.7,0.3) cosmology, we find $\lambda L_{\lambda}(1350\,{\rm \AA})\sim 3\times 10^{46}\,{\rm erg\,s^{-1}}$. } $\lambda L_{\lambda}(1350\,{\rm \AA})\sim 3\times 10^{46}\,{\rm erg\,s^{-1}}$. We focus on this object  as it has the best-sampled $B$ and $R$ light curves, as well as the highest fractional variability \citep[and references therein]{kas07}, $F_{\rm var}\simeq 0.16$, of a high-$z$ quasar in the \citet{geh03} sample. This object is consistent with being radio-quiet down to a flux limit of $\sim$\,6\,mJy~(5\,mJy) at 21\,cm (13\,cm) \citep{mar97,mau03}.  A more complete analysis of the MACHO quasar sample is deferred to a forthcoming publication.

\begin{figure}
\plotone{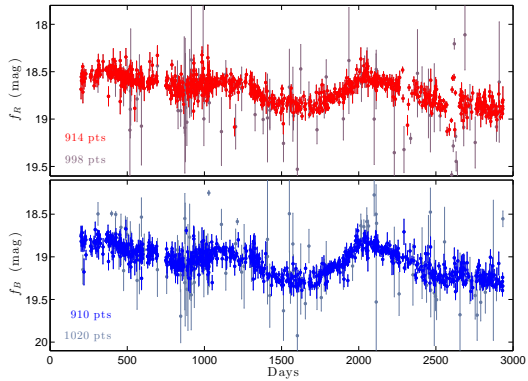}
\caption{$R$ (upper panel) and $B$ (lower panel) light curves for MACHO\,13.6805.324 cover $\sim 7.5$ years in the observed frame ($\sim 2.8$ years in the quasar frame). The number of points in each band is of order $10^3$, and depends on the filtering scheme used: darker shades correspond to effectively unfiltered data ($\zeta=10$) while brighter colors use a $\zeta=2$ filter. The fractional variability measure, $F_{\rm var}\simeq 0.14~(0.16)$ when $\zeta=2~(\zeta=10)$ filter is used. The mean photometric error is $\simeq 5.5\%~(7.6)\%$ for $\xi=2$ ($\xi=10$) .} 
\label{lc}
\end{figure}

We choose to work with the original MACHO bandpasses rather than transform into standard $B$ and $R$ bands.  This reduces the spectral overlap as the photometric transformations require linear combinations of the MACHO magnitudes. Light curves were filtered against poor data: all points with errors larger than $\zeta$ times the mean error or those which deviate from the light curve mean by more than $\zeta$ standard deviations were discarded (Fig. \ref{lc}).

\subsection{Spectral decomposition}

\begin{figure}
\plotone{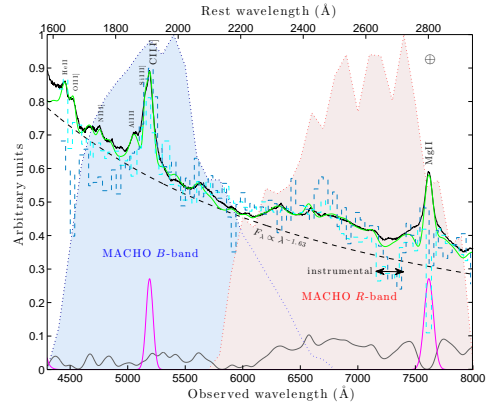}
\caption{A quasar composite spectrum at $z=1.72$ \citep{dvb01} overlaid with the throughput curves of the MACHO $B$- and $R$-filters [the somewhat erratic behavior is real and results from an interference pattern (C. Stubbs; private communication)]. A qualitative spectral decomposition into the prominent emission components is shown including the iron blend (gray curve), the \ion{C}{3}]\,$\lambda 1909$, and \ion{Mg}{2}\,$\lambda 2799$ lines (magenta curves), and a powerlaw continuum (dashed line). Designated minor emission lines were also included in the spectral decomposition (green curve). Integrating over the emission line flux, and taking into account the wavelength-dependent throughput, the largest contribution to line emission is from the iron complexes in the $R-$band. The contribution of the \ion{C}{3}\ and the \ion{Mg}{2}\ lines is roughly 4 times smaller than iron's and their relative contribution to their respective bands is  similar. Also shown are the two ({\it non} flux-calibrated, slightly shifted in wavelength range for clarity) optical spectra for MACHO\,13.6805.324, which were multiplied by a power-law function of wavelength so that their trend roughly follows that of the composite spectrum (dashed blue-shaded curves). The spectral shape is qualitatively consistent with the quasar template (note the instrumental feature at around 7300\AA) and suggests a significant contribution of the iron emission blend to the flux in the $R$-band. Note the prominent atmospheric absorption feature which happens to coincide with the \ion{Mg}{2}\ emission line at this particular redshift.} 
\label{filters}
\end{figure}

To effectively use the photometric RM technique of \citet{cd11}, it is helpful to identify the band with the larger contribution of emission lines to its variance. If the emission line contribution to the flux may be used as a proxy to its variable component (e.g., as would be the case if the relative flux variations in all emission lines were similar; see \S3), then by spectral decomposition and knowledge of the instrumental throughput, one can identify the line-rich and line-poor bands. Unfortunately, the quality of available spectra of MACHO\,13.6805.324 is low, and the data are not flux-calibrated. Nevertheless, their broad consistency with the composite quasar spectrum of \citet[see Fig. \ref{filters}]{dvb01} motivates us to use the latter for our purpose.

Prominent emission lines at $z=1.72$ include the \ion{C}{3}]\,$\lambda 1909$, \ion{Mg}{2}\,$\lambda 2799$, as well as the iron complexes (Fig. \ref{filters}). To account for the iron-blend contribution to the spectrum, we use the iron template of \citet{ves01}. An underlying powerlaw continuum model was assumed, $F_\lambda \propto \lambda^{-1.63}$, which is somewhat different from the one used by \citet{dvb01}, and better accounts for the flux level at $\sim 6000$\AA\ (observed), where little iron blend and Balmer continuum emission is present. Emission lines and blends were individually convolved with a single Gaussian kernel of some amplitude and width, and a qualitative agreement was sought with the \citet{dvb01} composite spectrum (Fig. \ref{filters}). The solution is not strictly unique, nor very physical, and its sole purpose is to {\it qualitatively} assess the relative contribution of the various emission components to the broadband flux. 

Weighing the decomposed spectrum by the instrumental throughput \citep{alc99}, we find that the $R$-band has the greater relative contribution of emission lines to its flux. Specifically, the iron emission blends contribute a net of $\sim 10$\% to the flux in the bands, with the relative contribution of the \ion{Mg}{2}\,$\lambda 2799$ and \ion{C}{3}]\,$\lambda 1909$ to their respective bands being $\gtrsim 4$ times smaller. 

\subsection{Uncovering the time-lag}

\begin{figure*}
\plotone{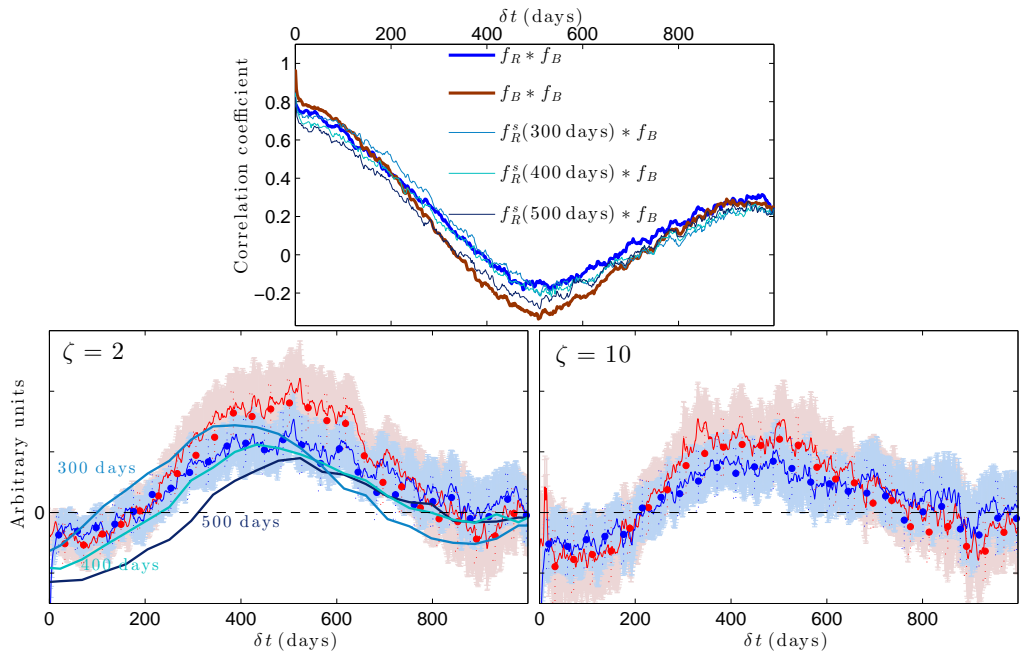}
\caption{Photometric RM analysis. {\it Top panel:} an excess power is seen at around 500\,days in the cross-correlation signal compared to the auto-correlation signal (thick lines). The simulated signal qualitatively reproduces the excess power in the cross-correlation term, at the right timescales, for three input time lags (see text and  legend). {\it Bottom panels:} the calculated $\xi$-functions for two levels of filtering: $\zeta=2$ (left) and $\zeta=10$ (right). Solid curves show the results of the ICCF algorithm with shaded regions indicating the uncertainty. The ZDCF calculations are shown as points with error bars in dotted lines. Blue (red) colors correspond to $\xi_{\rm CA}~(\xi_{\rm AA})$. All methods give consistent results and a modest level of light curve filtering yields a more significant result. We constrain the time scale at which the $\xi$-functions peak to be $\sim 500 \pm 100$\,days in the observed frame. Repeating the analysis using model-dependent synthetic $R$-band light curves (blue-shaded curves; see text) shows an agreement with a time lag of $\gtrsim$400\,days; considerably shorter or longer lags are excluded by the data.} 
\label{xis}
\end{figure*}

Having identified the line-rich band, we now proceed to measure the line-to-continuum time-delay. Following \citet{cd11}, we compute two statistical estimators for the line-to-continuum cross-correlation function: $\xi_{\rm CA}(\delta t)=f_R * f_B-f_B*f_B$, and  $\xi_{\rm AA}(\delta t)=f_R * f_R-f_B*f_B$ ($"*"$ denotes convolution and $f_B,~f_R$ are the light curves in the $B$ and $R$ bands, respectively; see Fig. \ref{xis}). With these definitions, a peak at $\delta t>0$ indicates an emission component in the $R$-band, which lags behind the $B$-band.  To calculate these statistical estimators, we use two independent schemes: the interpolated cross-correlation function \citep[ICCF]{pet04} and the $z$-transformed discrete correlation function \citep[ZDCF]{alx97}. The former has the advantage of being somewhat more sensitive, while the latter is less affected by sampling. 

Results for $\xi_{\rm CA}$ and $\xi_{\rm AA}$ using the ICCF and ZDCF schemes, and for two levels of light curve filtering, are presented in figure \ref{xis}. All analyses give consistent results: $\xi_{\rm CA}$ and $\xi_{\rm AA}$ peak at $\delta t \sim 400-600$\,days. Unsurprisingly, the peak is more significant when modest filtering ($\zeta=2$) is applied, resulting in $\xi_{\rm CA},~\xi_{\rm AA}$ being significant at the $\sim 2.8\sigma,~5\sigma$ ($\sigma$ is the standard deviation) level, respectively. 

To verify that the observed signal is indeed associated with a lagging component in the $R$-band, we reversed our choice of line-rich and line-poor bands, and repeated the analysis: no significant peak was detected at $\delta t>0$ (instead, a highly significant trough was detected at $\delta t \sim 400-600$\,days; not shown). In addition, we carried out sets of Monte Carlo simulations \citep{cd11} wherein the $B$-band light curve was convolved with a Gaussian kernel having prescribed time-delay and width (the latter was taken to be half of the former but the results are not very sensitive to the particular kernel chosen). The resulting light curve was then scaled down and added to the $B$-band signal so that the line contribution to the flux is $\sim 10\%$. The combined light curve was then sampled with the cadence of the $R$-band to create a synthetic light curve, $f_R^s$. The $\xi$-estimators were calculated for the light curve pair $[f_B,f_R^s]$, and the results for several input time-lags are shown in figure \ref{xis}. Clearly, the observed signal is qualitatively reproduced by our simulations for lags of the order of 400\,days. This a) confirms that the observed signal is indeed associated with a lagging emission component in the $R$-band and is not spurious, and b) that the time-lag constraints deduced below are meaningful and are not significantly biased. 

We estimate the time-lag and its uncertainty in the following way: the peak in $\xi_{\rm CA}$ and $\xi_{\rm AA}$ is identified with the time-lag for each Monte Carlo realization (a total of 100 realizations are used for each numerical scheme). A time-lag distribution is then obtained whose mean is identified with the time lag and its uncertainty with the standard deviation.  We find that the time-lag is $480\pm 100/530\pm70$\,days (ZDCF), or $490\pm 80/490\pm50$\,days (ICCF) for the $\xi_{\rm CA}/\xi_{\rm AA}$ statistical estimator. We conservatively estimate the lag to be $500\pm100$\,days, which corresponds to a rest-frame lag, $\tau\simeq180\pm40$\,days.

\section{Discussion \& Conclusions}

\begin{figure}
\plotone{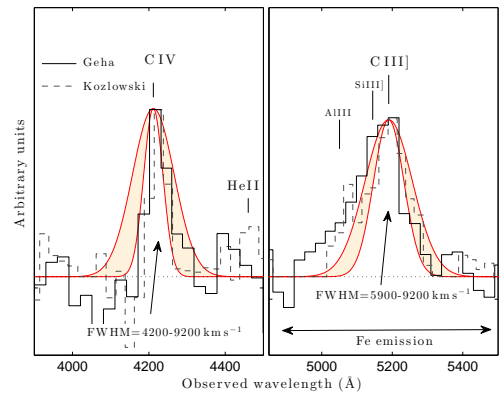}
\caption{The arbitrarily normalized (non flux-calibrated) spectra of MACHO\,13.6805.324 \citep[and S. Koz\l owski, private communication; see legend]{geh03} show two prominent emission lines. A cubic spline was fit far from the prominent line locations to normalize the spectrum (not shown), and single Gaussian models were overlaid to constrain the FWHM range, as indicated by the shaded areas. Only the red wing of the \ion{C}{4}\,$\lambda 1549$ line was used for fitting purposes due to potential absorption features just blue-ward of its peak (left panel). Similarly, only the red wing of the \ion{C}{3}]\,$\lambda 1909$ line was used for FWHM estimation to avoid blending with other emission lines just blueward of its center (right panel). We note that FWHM measurements for this line should be treated with caution since iron blend emission may be substantial (see also Fig. \ref{filters}), and is dififcult to estimate given the poor S/N of the non flux-calibrated data.} 
\label{fit}
\end{figure}

\begin{figure*}
\plottwo{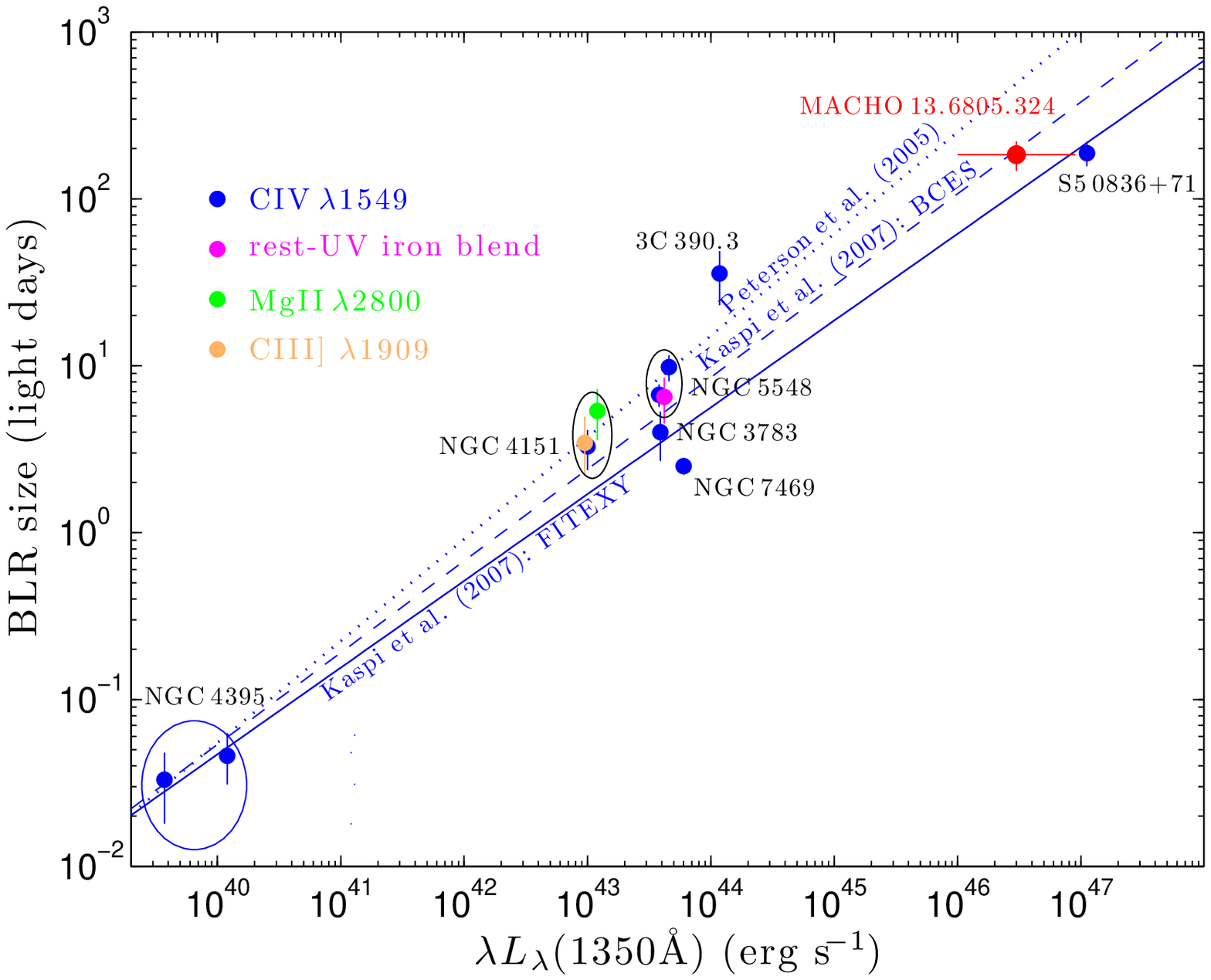}{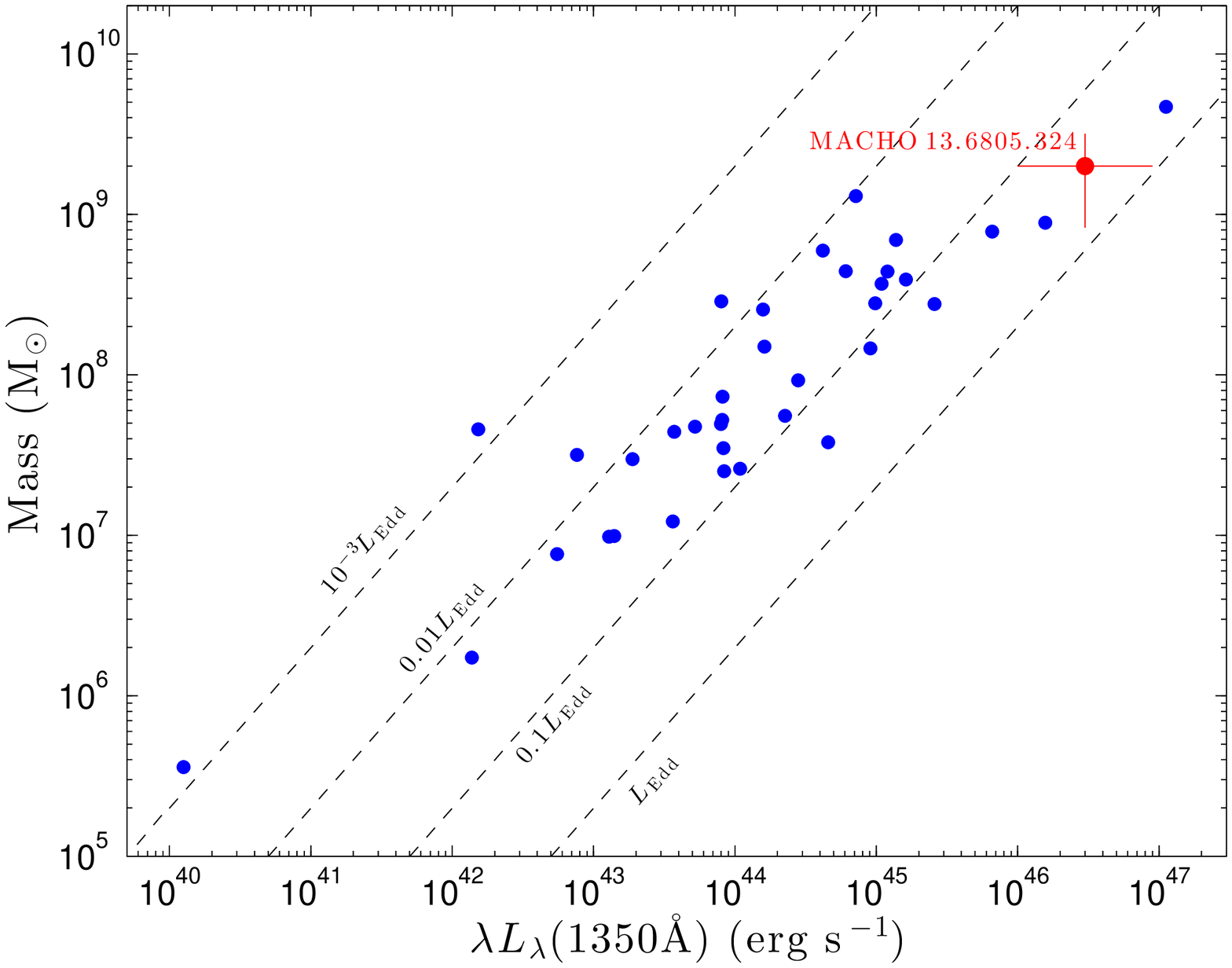}
\caption{The BLR size-luminosity relation (left; adopted from \citet{kas07}; see their paper for a discussion of the various fitted relations) and the mass-luminosity relation [right; adopted from \citet[and augmented by up-tp-date data from \citet{pet05,den06,kas07,gr08,ben09,den10,gr12}]{pet04} using the composite spectrum of \citet{dvb01} to convert monochromatic luminosities to $\lambda L_\lambda(1350\,{\rm \AA})$; uncertainties are not shown]. {\it Left:} plotted are data from \citet{pet04,pet05,kas05,kas07} as well as our results for MACHO\,13.6805.324, which are in agreement with both relations. Also shown, for comparison, are reliable BLR size measurements for the iron blend \citep{mao93} and the \ion{C}{3}\,$\lambda 1909$ and \ion{Mg}{2}\,$\lambda 2799$ emission lines \citep{met06}; see legend. Evidently, the BLR size measured here lies on the extrapolation of the $R-L$ relation, as obtained for low luminosity, low-$z$ objects. {\it Right:} Interestingly, MACHO\,13.6805.324 is consistent with emitting at $\gtrsim 10$\% of its Eddington rate, in agreement with the efficiency of other bright quasars [note that the mass reported here for S5\,0836+71 is $\sim$1.8 times larger than the value reported in \citet{kas07} and follows from the normalization of \citet{on04}]. Curves of constant $L/L_{\rm Edd.}$ are also plotted (dashed lines) assuming the bolometric correction used by \citet{kas07}.} 
\label{rel}
\end{figure*}

With only photometric data at hand, interpreting our findings is not straightforward: while we have a statistically robust time lag measurement, it is difficult to associate it with a specific emission line\footnote{Another possibility is that the signal originates in an unidentified continuum emission component.  While this cannot be excluded based on available data, we find it rather unlikely given that continuum emission, by definition, extends over a broad wavelength range, and the fact that there is no clear reason why the lag should be similar to that which characterizes emission lines.}. That being said, it is very likely that the observed time-lag is associated with the iron emission blends for the following reasons: 1) a positive peak in $\xi_{\rm CA},~\xi_{\rm AA}$ indicates that emission lines contributing to the $R$-band are responsible for the signal, 2) the iron contribution to the $R$-band typically exceeds that of the \ion{Mg}{2}\,$2799$\AA\ line by a factor $\ge 4$ (Fig. \ref{filters}), and 3) the implied flux variation of the iron blend is consistent with that seen in low-$z$ objects \citep{mao93,ves05}, and indirectly deduced for high-$z$ quasars \citep{meu11}\footnote{In a recent work, \citet{meu11} analyzed Stripe 82 SDSS quasars and found that the combined flux from all emission lines and blends varies by $\sim 10$\%, which is consistent with the relative variations measured for individual lines by \citet{kas07}. This implies that similar flux variations of the iron blend are characteristics of quasars [see also \citet{wil05} for some additional constraints on the time variability of emission lines in high-$z$ quasars].}.

Little is known with confidence about the physical properties of the UV iron-emitting region, and a statistically significant time-lag, reflecting on its size, exists for only one additional low-luminosity source, NGC5548 \citep{mao93}. Taken together, the results imply that the size-luminosity scaling for this region is consistent with that of other broad lines \citep[see our Fig. \ref{rel}]{kas05,kas07}. Interestingly, the inferred sizes of the iron- and  \ion{C}{4}\,$\lambda 1549$-emitting regions are comparable in NGC\,5548 \citep{mao93}. If true also for MACHO\,13.6805.324 then we find consistency with a simple extrapolation of size-luminosity relation for the \ion{C}{4}-emitting region, as obtained for lower luminosity sources, to the more luminous quasar population (Fig. \ref{rel}). A meaningful comparison with the size-luminosity relations for other rest-UV lines is currently limited by small number statistics.

Once the size of the line-emitting region is known, the black hole mass may be determined via $M_{\rm BH}\simeq 2.7\times 10^5({\rm FWHM}/10^3\,{\rm km\,s^{-1}})^2(\tau/{\rm days})\,{\rm M_\odot}$ \citep[and using the normalization of \citet{on04}]{pet04}, where FWHM is the full width at half maximum of the relevant emission line. This expression assumes that the gas is virialized and uses the FWHM as a proxy for its velocity dispersion in the black hole's potential well. Nevertheless, the FWHM of the iron blend in MACHO\,13.6805.324 cannot be determined from the available spectra, and it is not clear that the iron-emitting gas is virialized in quasars \citep{hu08}. With these uncertainties in mind, we note that currently available data point to a roughly  comparable size for the iron,  \ion{C}{4}\,$\lambda1549$, and \ion{C}{3}\,$\lambda 1909$ emitting regions \citep[see our Fig. \ref{rel}]{mao93,met06}\footnote{The situation concerning the size of the \ion{C}{3}\,$\lambda1909$ region is far from being clear: two low-luminosity objects having only loose constraints on the time-lag indicate that it may be twice the size of the \ion{C}{4}\,$\lambda 1549$ region \citep{pet94}, while a tentative time-lag measurement in a high-$z$ quasar indicates that the line traces the continuum level with no apparent delay \citep{kas07}.}. If true also for MACHO\,13.6805.324 then we may use the FWHM of the carbon lines and our iron line region size measurement to estimate $M_{\rm BH}$\footnote{There are uncertainties associated with the use of carbon lines for measuring $M_{\rm BH}$: as noted by \citet{bl05}, the \ion{C}{4}\,$\lambda 1549$ FWHM may be affected by non-virial motions, especially for bright objects emitting close to their Eddington limit. Also, line de-blending may be required to properly estimate the FWHM of the  \ion{C}{3}\,$\lambda 1909$ line. Lastly, a generic problem with single-epoch FWHM estimates concerns the fact that the FWHM of the varying component of the line may differ from that measured from the mean spectrum \citep{kas00}.}. Taking FWHM=$6500\pm2500\,{\rm km~s^{-1}}$ (Fig. \ref{fit}), we estimate the black hole mass to be $2\times 10^9\,{\rm M_\odot}$ with a formal uncertainty of $\sim 0.3$\,dex (see Fig. \ref{rel}). We note, however, that this mass estimate does not account for potentially important, but poorly understood, systematic effects concerning the physics and geometry of the iron-blend emitting gas.

It is worth noting that, being $>100$ quieter in the radio band than S5\,0836+71 \citep{kas07}, MACHO\,13.6805.324 is {\it not} a blazar, and may well be a radio-quiet object. As such, it may have the largest black hole mass ever measured using the RM technique in a representative member of the quasar population. Our findings also indicate that MACHO\,13.6805.324 is probably shining at $\gtrsim 10$\% of its Eddington luminosity (Fig. \ref{rel}), and in agreement with similar quasars in its class \citep{sh06}. 

A more reliable interpretation of our results, as well as a more accurate BH mass determination, require a higher signal-to-noise spectrum. A sparse (one visit every two months) spectroscopic-monitoring campaign of MACHO\,13.6805.324, over a period of $\sim7$ years, will determine the variability amplitude of the iron blend as well as potentially (independently) measure the time-lag in this object. Furthermore, by analyzing large enough samples of MACHO  quasars, with a similar redshift and luminosity range to MACHO\,13.6805.324, it may be possible to statistically corroborate the results presented here \citep{cd11}. With these future tests and improvements in mind, our present findings already demonstrate the feasibility of long-term photometric surveys in determining the BLR size (and BH mass) in luminous high-$z$ quasars.

\acknowledgements

We are grateful to Marla Geha for providing us with the MACHO quasar light curves and spectra in electronic form, and for commenting on an earlier version of this letter. We thank S. Koz\l owski for providing us with a recently acquired spectrum of MACHO\,13.6805.324, and  M. Vestergaard for supplying us with iron emission templates in electronic form. Fruitful discussions with H. Netzer and O. Shemmer are greatly appreciated, as well as helpful comments by the referee. This research has been supported in part by a FP7/IRG PIRG-GA-2009-256434 grant as well as by grant 927/11 from the Israeli Science Foundation.

\end{document}